\documentclass[twocolumn]{revtex4}
\usepackage{amssymb,amsmath,graphicx}

\newcommand{\bs}[1]{\boldsymbol{#1}}
\newcommand{\figref}[1]{Fig.~\ref{#1}}
\newcommand{\tabref}[1]{Table~\ref{#1}}
\newcommand{\refref}[1]{Ref.~\onlinecite{#1}}

\newcommand{\kB}{k_{\rm B}}

\newcommand{\K}{{\rm K}}
\newcommand{\fsigma}{{\bs{\sigma}}}

\begin{document}
\bibliographystyle{prsty}

\title{Cluster interactions for fcc-based structures in the Al-Mg-Si system} %90
\author{Nils Sandberg$^1$, Mattias Slabanja$^2$ and Randi Holmestad$^1$}
\affiliation{$^1$ Department of Physics, Norwegian University of Science and Technology (NTNU)\\
N-7491 Trondheim, Norway\\
$^2$ Department of Applied Physics, Chalmers University of Technology\\
S-412 96 Gothenburg, Sweden}
\date{\today}

% \twocolumn[\hsize\textwidth\columnwidth\hsize\csname
% @twocolumnfalse\endcsname

%%%%%%%%%%%%%%%%%%%%%%%%%%%%%%%%%%%%%%%%%%%%%%%%%%%%%%%%%
% ABSTRACT                                              %
%%%%%%%%%%%%%%%%%%%%%%%%%%%%%%%%%%%%%%%%%%%%%%%%%%%%%%%%%

\begin{abstract}

A class of proposed coherent precipitate structures 
(Guinier-Preston zones) in the Al-Mg-Si alloy are investigated using 
first-principles density functional theory methods. 
The cluster expansion method is used to extract effective 
interaction parameters, providing the means for large scale energy
calculations of alloy structures.
The Mg$_1$Si$_1$ L$1_0$ structure and structures related to the 
Mg$_5$Si$_6$ $\beta''$ phase are studied in more detail, 
and e.g., precipitate/matrix interface energies are presented. 
Using direct first-principles calculations we show that the former phase 
is dynamically unstable and thus must be stabilized by the surrounding 
Al matrix. 
Monte Carlo simulations and free-energy techniques are used to study the Al 
rich side of the phase diagram with the current CE parameters, and 
kinetic Monte Carlo simulations are used to study clustering in the 
disordered phase. 
The implications of our findings are discussed in the framework of 
classical nucleation theories, and we outline possible nucleation 
mechanisms.
\end{abstract}

\maketitle

%\pacs{PACS numbers: }
%\narrowtext
%\vspace*{0.5pc} ]

%%%%%%%%%%%%%%%%%%%%%%%%%%%%%%%%%%%%%%%%%%%%%%%%%%%%%%%%%%%%%%%%%%%%%%%%%%%%%%%%%%%%%%%%%
%%%%%%%%%%%%%%%%%%%%%%%%%%%%%%%%%%%%%%%%%%%%%%%%%%%%%%%%%%%%%%%%%%%%%%%%%%%%%%%%%%%%%%%%% 
%%%%%%%%% INTRODUCTION                                                         %%%%%%%%%%
%%%%%%%%%%%%%%%%%%%%%%%%%%%%%%%%%%%%%%%%%%%%%%%%%%%%%%%%%%%%%%%%%%%%%%%%%%%%%%%%%%%%%%%%%
%%%%%%%%%%%%%%%%%%%%%%%%%%%%%%%%%%%%%%%%%%%%%%%%%%%%%%%%%%%%%%%%%%%%%%%%%%%%%%%%%%%%%%%%%

\section{Introduction}
The formation of solute-rich precipitates in a super-saturated solid 
solution is a process of great technological importance because 
it gives raise to precipitation hardening in a wide class of alloys. 
Among the precipitation hardened Al alloys, the Al-Mg-Si (6xxx)
alloys belong to the most studied with regards to their precipitation 
sequences, see \refref{ZandAnde_sci97,EdwaStil98,MariAnde05} and references therein.
They often display a complex sequence of metastable phases forming under different
conditions (time, temperature and composition). The structures of almost all of these 
phases are known in detail from transmission electron microscopy (TEM) studies. 
However, such techniques give no direct information about the energetics 
of the phases of interest. 
 First-principles calculations can provide vital new information in 
this respect, as has recently been demonstrated in the case of the 
Al-Mg-Si system, see work by Derlet {\em et al}\cite{Derlet02} and Ravi and 
Wolverton.\cite{RaviWolv04} 
Thus, by means of solving the Schr{\"o}dinger equation within the density 
functional theory (DFT), one can obtain enthalpies, and study relaxation and 
binding, in phases which are meta-stable and only exist as nano-sized particles 
within the host matrix. 
A limitation of direct first-principles 
calculations is that, in practice, the structure has to be specified in advance. 
A central and unresolved question in the case of the Al-Mg-Si system is the structure 
of the first
precipitates that form from solid solution, i.e., the Guinier-Preston (GP) 
zones.\cite{MariAnde05}

In order to be able to study coherent precipitate phases within the Al matrix, 
and to model the disordered phase and the nucleation process, we have 
utilized a ternary cluster expansion (CE) description of the Al-Mg-Si system. 
In this model, a set of atoms within a lattice structure is denoted by 
$\fsigma$, and
the total energy of such a configuration can be written as a sum
of ``cluster functions'', $\Phi_{\alpha,s}(\fsigma)$;
\begin{equation}
  \label{eq:IsingSum}
E(\fsigma)=E_0+\sum_{\alpha,s} V_{\alpha,s}\Phi_{\alpha,s}(\fsigma).
\end{equation}
We derive the cluster interaction parameters $V_{\alpha,s}$ from a set 
of first-principles calculations using an approach based on the structure inversion 
method (SIM).\cite{ConnWill83}
Once these parameters have been obtained, it is a relatively
inexpensive operation to sum up the total energy. 
A complication that arises is that the effective cluster interactions 
are in principle volume dependent,\cite{LaksFerr92}
but since the concentration of Mg and Si in typical Al-Mg-Si alloys is $1-2$\%, 
we have done all calculations at volumes corresponding to the Al equilibrium 
lattice constant.

We have focused on the two main candidates for early precipitate 
structures, or GP-zones, in the Al-Mg-Si system. 
These are, the Mg$_1$Si$_1$ L$1_0$ structure proposed by Matsuda {\em et al},\cite{Matsuda98} 
here referred to as  the Matsuda structure, and the so called pre-$\beta''$-phase which 
was proposed by Marioara {\em et al}.\cite{MariAnde01}
By testing a large number of 
structures related to the pre-$\beta''$ structure, we conclude 
that the Matsuda-phase is the energetically most stable precipitate phase.
We then proceed with studying the stability of the Matsuda and pre-$\beta''$ 
structures and find that the former structure is dynamically unstable, 
so that the vibrational free energy is undefined. 
This implies that it can only exist within a surrounding stabilizing 
lattice, the Al lattice in this case. That this system exhibits instabilities 
is not surprising since fcc Si is dynamically unstable.\cite{EkmaPers00} 
For certain local concentrations of Si we therefore expect the lattice to be 
significantly softened, which lowers the Gibbs free energy, and may lead 
to structural transformations. 

The CE method also allows for large-scale Monte Carlo (MC) simulations to be performed,
either for finding ground-state structures (using, e.g., simulated annealing) or to study
finite-temperature thermodynamic properties of an alloy. Our MC calculations using the 
present CE parameters 
show that decomposition typically occurs in the form of spherical precipitates with the 
Matsuda structure. 
MC simulations in combination with thermodynamic integration are used to calculate the 
solvus phase boundary in the Al rich end of the phase diagram. 
We also use kinetic Monte Carlo (kMC) simulations to study clustering in the disordered phase. 
Such studies are of current interest because modern experimental techniques, i.e., high-
resolution transmision electron microscopy and 3D atom probe tomography, are capable of studying 
small, coherent clusters at an atomic scale.\cite{MariAnde05,ShaCere05} Therefore, increased 
understanding of atomic 
level processes that preceed nucleation can be expected and a much more detailed comparison with 
modelling and simulation is possible.\cite{ShaCere05}

The paper is disposed as follows. In Sec. II we describe the two GP-zone models 
considered in this work. In Sec. III we review the theory of 
ternary CE, and we also detail the first-principles calculations that form 
the basis for the present work. The results of the fitting procedure, and 
of the CE calculations for different GP-zone structures are presented in 
section IV. The calculation of the solvus boundary for Mg$_1$Si$_1$ and simulations of the 
early stage of clustering are presented in the same section.
 In Sec. V we discuss the present results in the light of what is 
experimentally known about GP-zone formation in Al-Mg-Si, and conclusions follow 
in Sec.~VI.

%%%%%%%%%%%%%%%%%%%%%%%%%%%%%%%%%%%%%%%%%%%%%%%%%%%%%%%%%%%%%%%%%%%%%%%%%%%%%%%%%%%%%%%%%
%%%%%%%%%%%%%%%%%%%%%%%%%%%%%%%%%%%%%%%%%%%%%%%%%%%%%%%%%%%%%%%%%%%%%%%%%%%%%%%%%%%%%%%%% 
%%%%%%%%% GP-zone models in the Al-Mg-Si alloys                               %%%%%%%%%%%
%%%%%%%%%%%%%%%%%%%%%%%%%%%%%%%%%%%%%%%%%%%%%%%%%%%%%%%%%%%%%%%%%%%%%%%%%%%%%%%%%%%%%%%%%
%%%%%%%%%%%%%%%%%%%%%%%%%%%%%%%%%%%%%%%%%%%%%%%%%%%%%%%%%%%%%%%%%%%%%%%%%%%%%%%%%%%%%%%%%
\section{GP-zone models in the Al-Mg-Si alloys}
\begin{figure}
  \begin{center}
  \includegraphics*[width=4cm]{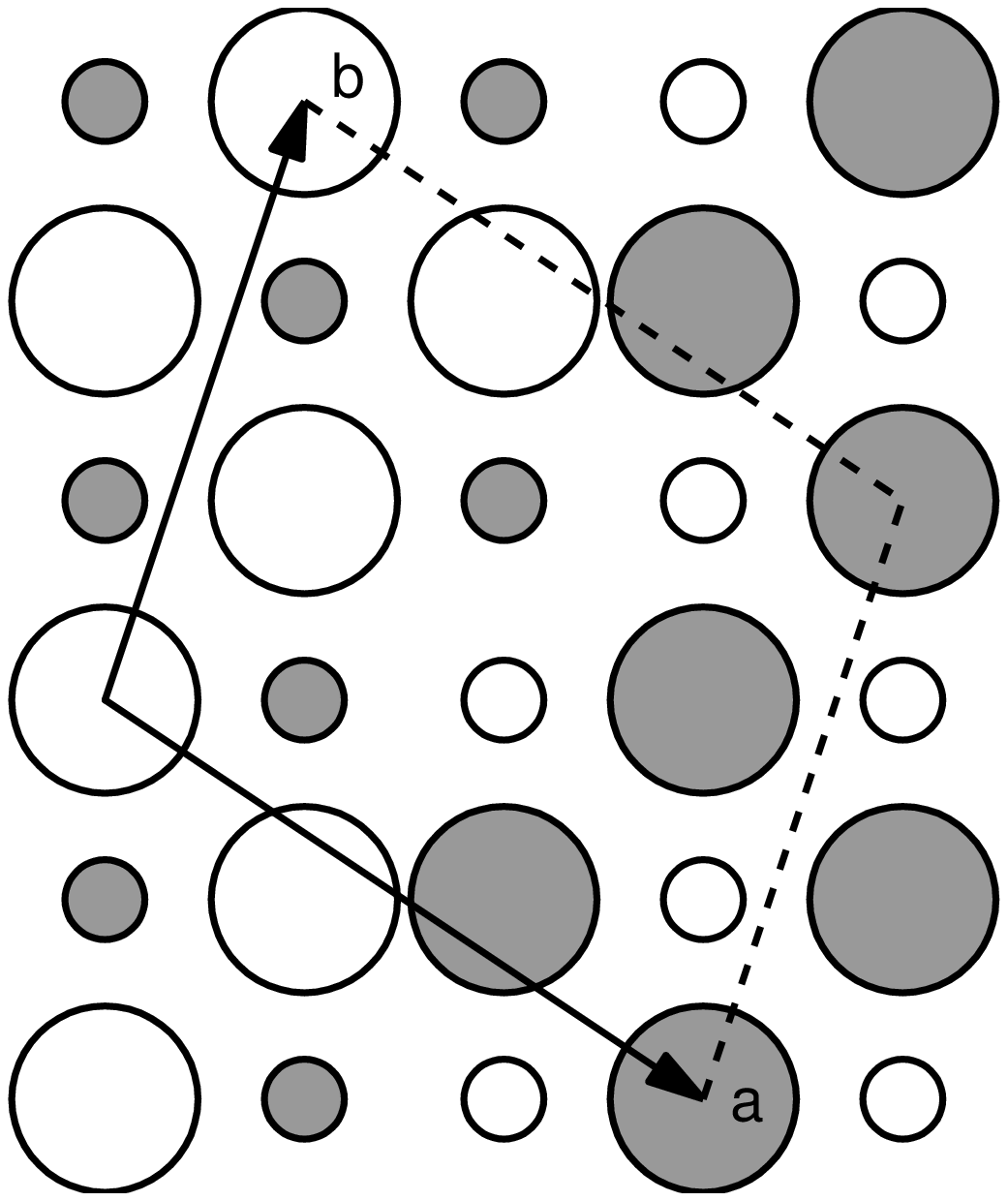}
  \includegraphics*[width=3cm]{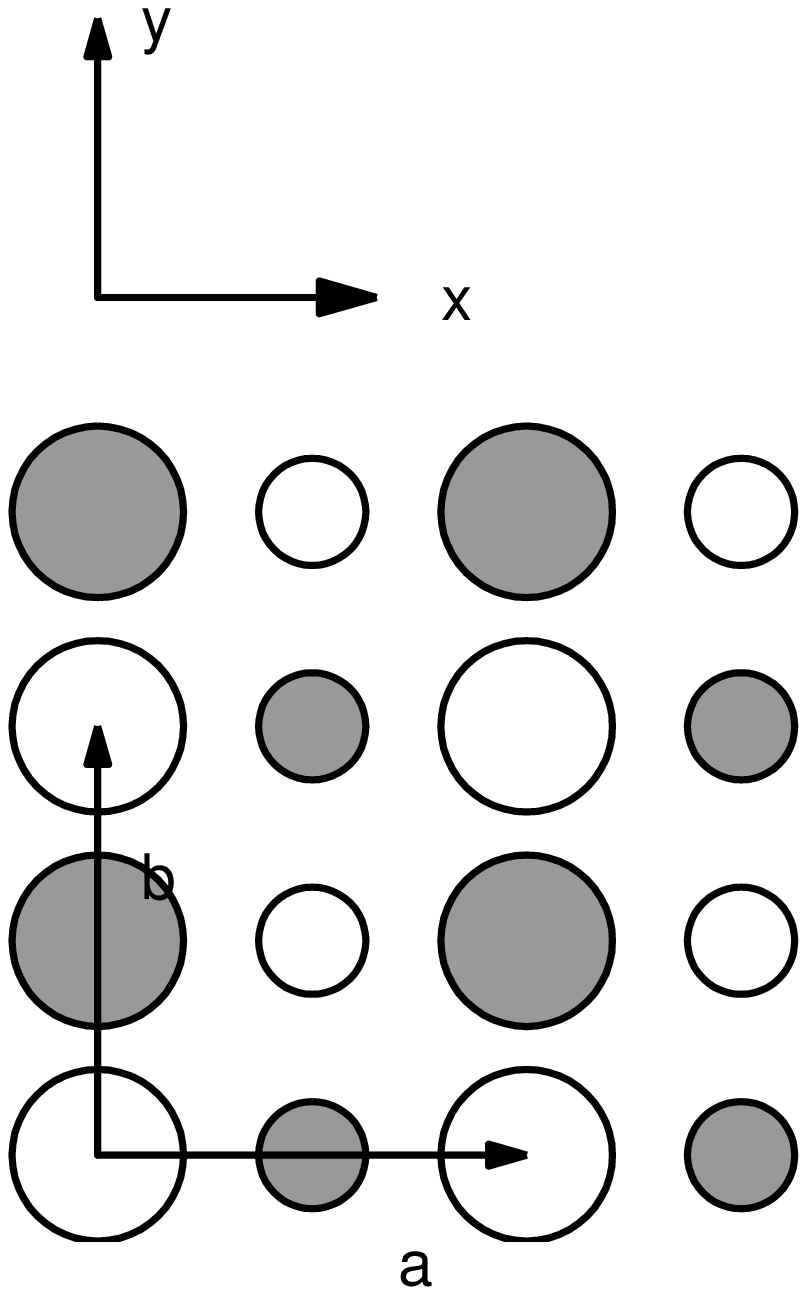}
  \caption{Previously proposed GP-zone structures, where large (small) 
           circles indicate Mg (Si) atoms. Dark atoms are displaced by 
           $1/2\;a_0$ in the $z$-direction, with respect to the white atoms.
           a) the pre-$\beta''$ structure 
           with composition Mg$_5$Si$_6$, the 11 atom primitive unit cell 
           is indicated. b) the Matsuda L$1_0$ structure with composition 
           Mg$_1$Si$_1$.}
           \label{fig:GP}
  \end{center}
\end{figure}

Currently there are two generally accepted models of the early precipitate phases, or 
GP-zones, in the Al-Mg-Si alloys.
The first structure, an Mg$_1$Si$_1$ L$1_0$ phase (see Fig \ref{fig:GP}b) was observed 
by Matsuda and co-workers~\cite{Matsuda98} in samples with a solute concentration of 
1.55 at.\% and with a Mg:Si 
ratio of 2:1. The samples had been aged and then heat-treated at 343 K. This phase will 
be referred to as the Matsuda phase.
The other structure, the pre-$\beta''$ in Fig.~\ref{fig:GP}a was 
proposed by Marioara et al.~\cite{MariAnde01} based on TEM observations of samples that had
been homogenized at 843 K and heat treated at 450 K. The Mg:Si ratio in these samples was 5:6. 

The $\beta''$ structure is the most efficient hardening phase in the 6xxx alloys. 
 The fact that GP-zones have been observed to have the same
symmetry as the $\beta''$ phase, motivated the structural model of GP-zones shown in
\figref{fig:GP}a.\cite{MariAnde01}
Its structural relationship with the $\beta''$ phase can
be understood in the following way.
If the $(0,0,0)$ Mg atom in 
Fig.~\ref{fig:GP}a is displaced $1/2a_0$ in the normal ($z$) direction, one obtains a
structural model of the $\beta''$ phase. The latter phase is in turn needle-shaped, and  often 
observed to be coherent in the $(x\;y)$ plane and semi-coherent in the $z$ (needle) direction.
%%%%%%%%%%%%%%%%%%%%%%%%%%%%%% FIGURE 1 %%%%%%%%%%%%%%%%%%%%%%%%%%%%%%%%%%%%%%%%%%%%%%%%%%%%
%%%%%%%%%%%%%%%%%%%%%%%%%%%%%% end FIGURE %%%%%%%%%%%%%%%%%%%%%%%%%%%%%%%%%%%%%%%%%%%%%%%%%%%

%%%%%%%%%%%%%%%%%%%%%%%%%%%%%%%%%%%%%%%%%%%%%%%%%%%%%%%%%%%%%%%%%%%%%%%%%%%%%%%%%%%%%%%%%
%%%%%%%%%%%%%%%%%%%%%%%%%%%%%%%%%%%%%%%%%%%%%%%%%%%%%%%%%%%%%%%%%%%%%%%%%%%%%%%%%%%%%%%%% 
%%%%%%%%% METHOD        	                                              %%%%%%%%%%%
%%%%%%%%%%%%%%%%%%%%%%%%%%%%%%%%%%%%%%%%%%%%%%%%%%%%%%%%%%%%%%%%%%%%%%%%%%%%%%%%%%%%%%%%%
%%%%%%%%%%%%%%%%%%%%%%%%%%%%%%%%%%%%%%%%%%%%%%%%%%%%%%%%%%%%%%%%%%%%%%%%%%%%%%%%%%%%%%%%%
\section{Methods}

\subsection{Cluster expansion and the structure inversion method}
The general theory for multicomponent CE was developed by Sanches, 
Ducastelle and Gratias.\cite{SancDuca84} 
In the present work we use the following notations. 
A lattice site $i$ can be occupied by an atom of type $A$, $B$, or $C$,
indicated by an occupation variable, $\sigma_i$, taking the value
$-1$, 0, or $+1$, respectively. 
Hence, given an underlying lattice, the alloy structure can be 
defined by a set of occupation variables, $\fsigma=\{\sigma_i\}$.
Linearly independent basis vectors describing the occupation are then 
\begin{eqnarray}
\label{eq:ClBasis}
\Theta^{(0)}_i &=& 1,\nonumber\\ 
\Theta^{(1)}_i &=& \sqrt{3/2}\sigma_i,\\
\Theta^{(2)}_i &=& \sqrt{2}(1-3\sigma_i^2)\nonumber. 
\end{eqnarray}
A cluster $\alpha$ is defined as a set of lattice sites $(p_1,p_2,...,p_n)$.
Given a cluster $\alpha$, a cluster function is defined as the the product
of the basis vectors,
\begin{equation}
  \label{eq:ClFunc}
  \Phi_{\alpha,s}(\fsigma)=\Theta^{(s_1)}_{p_1}\Theta^{(s_2)}_{p_2},...
\end{equation}
where the sequence $s=(s_1,s_2,...,s_n)$ is called ``decoration'', which in
the ternary case will consist of 1:s and 2:s.
The total energy is given by \eqref{eq:IsingSum}, 
and by considering the average of cluster functions for 
symmetry-equivalent cluster-decoration pairs, $\bar{\Phi}_{\alpha,s}(\fsigma)$,
the total energy can be re-written as
\begin{equation}
  \label{eq:CEenergy}
E(\fsigma)=E_0+\sum_{\alpha,s} V_{\alpha,s}m_{\alpha,s}\bar{\Phi}_{\alpha,s}(\fsigma),
\end{equation}
where $E_0$ is a reference energy and $\bar{\Phi}_{\alpha,s}(\fsigma)$ is an average, defined as
\begin{equation}
  \label{eq:ClMean}
  \bar\Phi_{\alpha,s}(\fsigma)=\frac{1}{N}\sum_{\alpha',s'} \Phi_{\alpha',s'}(\fsigma).
\end{equation}
The sum goes over all $(\alpha',s')$ pairs related to $(\alpha,s)$
through lattice symmetry operations. $m_{\alpha,s}$ is the number of such cluster functions
per lattice site, and $N$ is the total number of lattice sites. 

In the structure inversion method one calculates the $V_\alpha$:s by minimizing
\begin{equation}
  \label{eq:CE-fit}
\omega=\frac{1}{n}\sum_{\{\fsigma\}}
   [E_{\rm DFT}(\fsigma)-\sum m_{\alpha,s}V_{\alpha,s}\hat{\Phi}_{\alpha,s}(\fsigma)]^2
\end{equation}
where $\{\fsigma\}$ denotes a list of $n$ input structures and
$E_{\rm DFT}(\fsigma)$ their corresponding total energies.
The choice of input structures is arbitrary as long as they contain 
information about all cluster functions, and an advantage with the 
SIM is that the $\fsigma$:s can consist of small unit
cells in a super-cell approach, for which the total energy, 
may be efficiently calculated using DFT methods.

In order to select cluster functions to be used in the fitting, 
we used cross-validation,\cite{WallCede02} with the intention 
to obtain an optimal fit, while at the same time avoid over-fitting.
It is applied to CE by defining a hierarchy of clusters, were a cluster
is included only if all its sub-clusters are included,
and short range clusters are prioritized over long-range clusters.
Then, by choosing clusters in this way, the aim is to minimize the
cross-validation score
\begin{equation}
  \label{eq:cv-score}
(cv)^2=\frac{1}{n}\sum_{i=1}^n (E_i-E_{(i)})^2.
\end{equation}
Here, $E_{(i)}$ denotes the predicted energy of structure $i$, 
using parameters $V_{\alpha,s}$ obtained {\em without} structure $i$
included in the fit. Thus, the cv-score measures the ability of 
the CE, with a given set of cluster functions, to predict the 
energy of new structures. We note that all $n$ structures are 
included in the fit in Eq.~\eqref{eq:CE-fit}.  

%%%%%%%%%%%%%%%%%%%%%%%%%%%%%%%%%%%%%%%%%%%%%%%%%%%%%%%%%%%%%%%%%%%%%%%%%%%%%%%%%%%%%%%%%%%%%%
%%%%%%%%%%%%%%%%%%%%%%%%%%%%%%%%%%%%%%%%%%%%%%%%%%%%%%%%%%%%%%%%%%%%%%%%%%%%%%%%%%%%%%%%%%%%%%
%%%%%%%%% ELECTRONIC STRUCTURE CALCULATIONS	                                   %%%%%%%%%%%
%%%%%%%%%%%%%%%%%%%%%%%%%%%%%%%%%%%%%%%%%%%%%%%%%%%%%%%%%%%%%%%%%%%%%%%%%%%%%%%%%%%%%%%%%%%%%%
%%%%%%%%%%%%%%%%%%%%%%%%%%%%%%%%%%%%%%%%%%%%%%%%%%%%%%%%%%%%%%%%%%%%%%%%%%%%%%%%%%%%%%%%%%%%%%
\subsection{Electronic structure calculations}

We have used the Vienna ab-initio simulation package (VASP),\cite{VASP} a plane-wave pseudo 
potential implementation of the DFT method, in the present 
electronic structure calculations. 
For the exchange-correlation approximation we used the Perdew-Wang implementation 
of the generalized gradient approximation (GGA).\cite{PW91} 
The built-in ultra-soft pseudo potentials by Vanderbilt {\it et al.} were used.\cite{PP}
$k$-points were chosen according to the Monkhorst-Pack scheme in meshes 
 corresponding to $12\times12\times12$ 
points for the conventional fcc unit cell, or denser. 
Tests showed that using a mesh corresponding to $16\times16\times16$ $k$-points changed 
the energy with typically 1-2 meV/atom.
The atomic structure was relaxed at constant volume using a conjugate gradient 
technique, or in some cases a damped molecular dynamics algorithm.
We found that both methods led to the same relaxed state. 
We note that during the relaxation, the symmetry is preserved. 
This means that if the relaxation starts at a saddle point in the potential
energy landscape, the system may stay at this saddle point due to symmetry 
restrictions.
In particular, pure Si in the fcc structure is unstable with respect to 
tetragonal shear,\cite{EkmaPers00} but this relaxation mode is not taken into account in our 
calculations. 

%%%%%%%%%%%%%%%%%%%%%%%%%%%%%%%%%%%%%%%%%%%%%%%%%%%%%%%%%%%%%%%%%%%%%%%%%%%%%%%%%%%%%%%%%%%%%%
%%%%%%%%%%%%%%%%%%%%%%%%%%%%%%%%%%%%%%%%%%%%%%%%%%%%%%%%%%%%%%%%%%%%%%%%%%%%%%%%%%%%%%%%%%%%%%
%%%%%%%%% MONTE CARLO ETC THEORY              	                                   %%%%%%%%%%%
%%%%%%%%%%%%%%%%%%%%%%%%%%%%%%%%%%%%%%%%%%%%%%%%%%%%%%%%%%%%%%%%%%%%%%%%%%%%%%%%%%%%%%%%%%%%%%
%%%%%%%%%%%%%%%%%%%%%%%%%%%%%%%%%%%%%%%%%%%%%%%%%%%%%%%%%%%%%%%%%%%%%%%%%%%%%%%%%%%%%%%%%%%%%%
\subsection{Monte Carlo and kinetic Monte Carlo simulations}

The computational study of the thermodynamic properties of the
Al-Mg-Si system requires to take configurational space properly into
account.
To this end we use the standard Metropolis\cite{metropolis53} Monte Carlo
sampling technique, generating a sequence of configurations being 
Boltzmann-distributed with respect to the system temperature and
configurational energy. 
We have held the alloy constituent
concentrations unchanged during the simulations.
Hence, as the trial-changes we simply choose a random pair of atoms
to be exchanged with each-other. One MC step is defined as one such trial-echange 
per atom.
A technical aspect of our implementation is that the cluster functions are defined 
in the so called canonical basis
rather than the orthogonal basis (Eq:s \eqref{eq:ClBasis}). That means that the cluster
functions are products of integers rather than real numbers, and their 
evaluation becomes more efficient.

In order to simulate kinetic processes (e.g., nucleation)
we have utilized a kinetic Monte Carlo (kMC) algorithm.
The important additional constraint added to a kMC algorithm, as
compared to the ordinary MC dito, is that the (trial-) changes made to
the system are selected in a way correspondig to the effect of some
physical process occurring.
The physical process being most important during aging or phase
transformations in our Al-Mg-Si alloy system, is the mass transport 
via atomic diffusion.
The atomic diffusion process was modelled by attempted
interchange of two randomly choosen nearest neighbor atoms, and accepting
the interchange according to the Metropolis algorithm. 
This model is also known as Kawasaki-dynamics.\cite{kawasaki72}

To provide an estimate of the time-scale of the simulated
diffusion process, we use an experimental value of the aluminum self diffusion
coefficient\cite{hatch84} as input.
The experimental input which implicitly contain information about diffusion
hop-rates and vacancy concentrations, will relate the Monte Carlo
steps to physical time as
\begin{equation}
  \tau(T) \sim \frac{a_{\rm 0}^{2}}{D_{\rm exp}(T)}\frac{1}{s_{\rm v}},
\end{equation}
where $\tau$ is the physical time scale of giving all the atoms in the system a
chance to diffuse, $a_{\rm 0}$ being the Al lattice contant, and 
$D_{\rm exp}(T)$ is the experimentally determined diffusion coefficient.
The factor $s_{\rm v}$ accounts for the supersaturation of vacancies 
that is present, e.g., after a quench.

We used the kMC method to study the early stage of clustering in Al-Mg-Si. 
Our definition
of a cluster is then that each solute atom (Mg or Si) belonging to a given cluster is 
nearest neighbour to at least one more solute atom in the same cluster.

%%%%%%%%%%%%%%%%%%%%%%%%%%%%%%%%%%%%%%%%%%%%%%%%%%%%%%%%%%%%%%%%%%%%%%%%%%%%%%%%%%%%%%%%%%%%%%
%%%%%%%%%%%%%%%%%%%%%%%%%%%%%%%%%%%%%%%%%%%%%%%%%%%%%%%%%%%%%%%%%%%%%%%%%%%%%%%%%%%%%%%%%%%%%%
%%%%%%%%% FREE ENERGY ETC THEORY              	                                   %%%%%%%%%%%
%%%%%%%%%%%%%%%%%%%%%%%%%%%%%%%%%%%%%%%%%%%%%%%%%%%%%%%%%%%%%%%%%%%%%%%%%%%%%%%%%%%%%%%%%%%%%%
%%%%%%%%%%%%%%%%%%%%%%%%%%%%%%%%%%%%%%%%%%%%%%%%%%%%%%%%%%%%%%%%%%%%%%%%%%%%%%%%%%%%%%%%%%%%%%
\subsection{Free energy integration}
The phase diagram, and the underlying free-energy differences between 
various phases determine the driving force, e.g., for precipitate formation. 
In order to calculate the phase boundary between disoerdered and a two-phase 
region, one must know the free energy of the disordered phase. 
In the present work, it was calculated in the following way. 
The energy as a function of temperature was determined for a set of 
compositions in standard MC simulations. This was done by slowly decreasing the 
temperature and the system was then kept in the disorderd state. A corresponding 
simulation where the temperature is increased from a low temperature shows a 
clear hysteresis and the point where the two curves meet can be taken as a 
rough estimate of the phase boundary. The entropy in the disordered state was 
determined using thermodynamic integration of the relation
\begin{equation}
\label{eq:ds_dh}
\frac{dS}{dT}=\frac{1}{T}\frac{dH}{dT}.
\end{equation}
The remaining constant of intgration was taken from the expression for 
the entropy of a ternary system in the ideal solution model
\begin{equation}
\label{eq:S_HT}
S_{\rm ideal} = -\kB(c_1\ln(c_1)+c_2\ln(c_2)+c_3\ln(c_3))
\end{equation}
where $c_1$, $c_2$ and $c_3$ are the respective atomic concentrations.
This expression is which is valid at sufficiently high temperature. In the present work, 
we integrated from $T=3000\;\K$. 

%%%%%%%%%%%%%%%%%%%%%%%%%%%%%%%%%%%%%%%%%%%%%%%%%%%%%%%%%%%%%%%%%%%%%%%%%%%%%%%%%%%%%%%%
%%%%%%%%%%%%%%%%%%%%%%%%%%%%%%%%%%%%%%%%%%%%%%%%%%%%%%%%%%%%%%%%%%%%%%%%%%%%%%%%%%%%%%%% 
%%%%%%%%%                         RESULTS                                    %%%%%%%%%%%
%%%%%%%%%%%%%%%%%%%%%%%%%%%%%%%%%%%%%%%%%%%%%%%%%%%%%%%%%%%%%%%%%%%%%%%%%%%%%%%%%%%%%%%%
%%%%%%%%%%%%%%%%%%%%%%%%%%%%%%%%%%%%%%%%%%%%%%%%%%%%%%%%%%%%%%%%%%%%%%%%%%%%%%%%%%%%%%%%
\section{Results}

\subsection{Fitting of the CE parameters}

To choose input structures for the SIM, we started with 8 binary structures  
and 34 ternary structures in Refs. \onlinecite{Ducastel91} and 
\onlinecite{CedeGarb94}, respectively. 
Those structures have been shown to be the only distinct structures that can 
be obtained if one assumes only up to next-nearest neighbor pair interactions. 
(In the ternary case, this has not been strictly proven, but it seems plausible.) 
Taking into account the structures obtained when the occupation of A, B or C 
atoms in the 'generic' structures are permuted, one ends up with 148 
symmetrically distinct structures. In addition, we used some low-concentration 
structures, e.g., an isolated impurity in a 32 LP cell, see Appendix A. 
Further, we excluded some structures, e.g., pure Mg, that had a large static 
(positive or negative) internal pressure. 
In total, we ended up with 154 structures. 

\begin{figure}
  \begin{center}
    %% {data/cvScore/cvScore.eps}
    \includegraphics*[width=7cm]{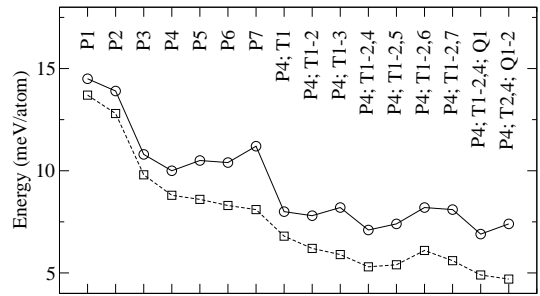}
    \caption{\label{fig:Conv}
             Convergence of the cv-score (circles) and the mean-square deviation 
             between predicted and calculated energies
             of the input structures (squares), as a function of included clusters; 
             pairs (P), three-point clusters (T), and four-point clusters (Q). 
             It is understood that, e.g., P4 means P1-4. See further Table 1.}
  \end{center}
\end{figure}

These structures were used as input to the fitting of the cluster interaction 
parameters in Eq.~\eqref{eq:CE-fit}. 
When selecting clusters, we minimized the cross-validation score given by 
Eq.~\eqref{eq:cv-score}. 
In the procedure we included pairs, triplets and quadruplets in a hierarchical 
order, see \figref{fig:Conv} and Table \ref{tb:1} for an explanation of the 
clusters used. 
We found that a good fit was obtained if we included pairs up to 
fourth nearest neighbor, and the triplets T1, T2 and T4 (see \tabref{tb:1}). 
This gives a total of $27$ parameters $V_{\alpha,s}$ that were minimized. 
The resulting cv-score is $cv=$~7.1~meV/atom.

%%%%%%%%%%%%%%%%%%%%%%%%%%%%%% FIGURE %%%%%%%%%%%%%%%%%%%%%%%%%%%%%%%%%%%%%%%%%%%%%%%%%%%%%%%%%
%%%%%%%%%%%%%%%%%%%%%%%%%%%%%% end FIGURE %%%%%%%%%%%%%%%%%%%%%%%%%%%%%%%%%%%%%%%%%%%%%%%%%%%%%

%%%%%%%%%%%%%%%%%%%%%%%%%%%%%%%%%%%%%%%%%%%%%%%%%%%%%%%%%%%%%%%%%%%%%%%%%%%%%%%%%%%%%%%%
%%%%%%%%%%%%%%%%%%%%%%%%%%%%%%%%%%%%%%%%%%%%%%%%%%%%%%%%%%%%%%%%%%%%%%%%%%%%%%%%%%%%%%%% 
%%%%%%%%%                        STATIC ENERGIES                             %%%%%%%%%%%
%%%%%%%%%%%%%%%%%%%%%%%%%%%%%%%%%%%%%%%%%%%%%%%%%%%%%%%%%%%%%%%%%%%%%%%%%%%%%%%%%%%%%%%%
%%%%%%%%%%%%%%%%%%%%%%%%%%%%%%%%%%%%%%%%%%%%%%%%%%%%%%%%%%%%%%%%%%%%%%%%%%%%%%%%%%%%%%%%
\subsection{Static energies of suggested GP-zone structures}
We have investigated the energetics of two models of the GP-zones in Al-Mg-Si;
the Matsuda phase and the pre-$\beta''$ phase (see Sec.~II and \figref{fig:GP}).

\begin{figure}
  \begin{center}
  %% {data/toten1.1/E.eps}
  \includegraphics*[width=8cm]{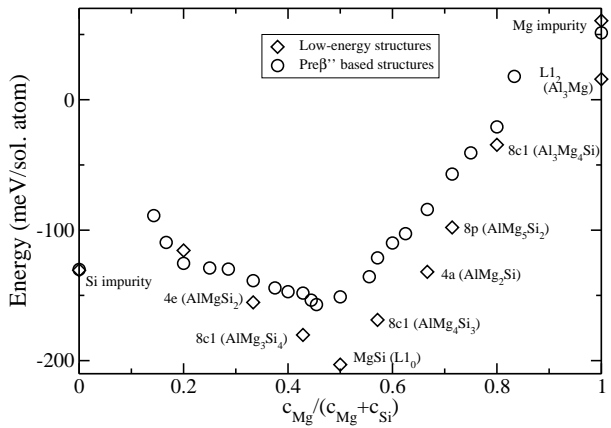}
  \caption{Total energy per solute atom of different suggested GP-zone structures, along 
           with some of the structures used in the fitting procedure. 
           The results are given with reference to the energies of the pure elements, 
           in the fcc structure, at zero pressure. 
	   The structures marked with diamonds are 
           denoted in accordance with~\protect\refref{CedeGarb94}.}
  \label{fig:toten1}
  \end{center}
\end{figure}

Because the latter structure has not actually been solved, we proceed along 
the lines in \refref{RaviWolv04}: starting with the 11 atom unit cell in 
\figref{fig:GP}a, we exchanged Mg/Si atoms with Al atoms in all $2^{11}=2048$ 
possible ways. 
For each structure (of which some will be identical by symmetry) we calculated 
the total energy using Eq.~\eqref{eq:IsingSum}. In Fig.~\ref{fig:toten1} we have plotted, 
for each 
composition $c_{\rm Mg}/(c_{\rm Mg}+c_{\rm Si})$, the energy of the most stable
structure.
An even less restrictive procedure is to generate all $3^{11}=177147$ structures 
that arise if each lattice point is occupied by a Al, Mg or Si atom. 
The most stable structures in this case are essentially the same as those in 
\figref{fig:toten1}.

%%%%%%%%%%%%%%%%%%%%%%%%%%%%%% FIGURE 3 %%%%%%%%%%%%%%%%%%%%%%%%%%%%%%%%%%%%%%%%%%%%%%%%%%%%%%%
%%%%%%%%%%%%%%%%%%%%%%%%%%%%%% end FIGURE 3 %%%%%%%%%%%%%%%%%%%%%%%%%%%%%%%%%%%%%%%%%%%%%%%%%%%%%%%

Ravi and Wolverton\cite{RaviWolv04} did a set of calculations for pre-$\beta''$ 
structures, using first-principles methods. 
We have used the same convention as they used in plotting the energies, i.e., we plot
the formation energies per solute atom with respect to the pure constituents in the
fcc structure. Therefore, the the two sets of energies can be directly compared. 

\begin{figure}
  \begin{center}
  %% {data/toten2/E.eps}
  \includegraphics*[width=7cm]{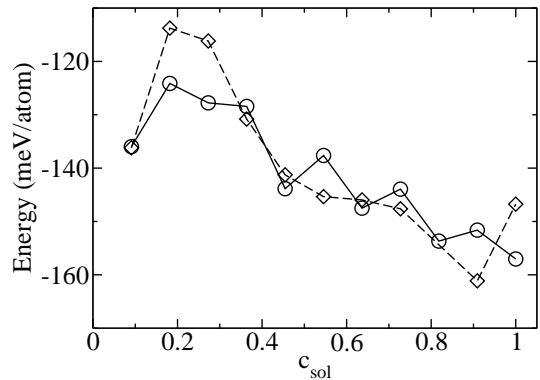}
  \caption{Predicted ground state structures in the 11 atom unit cell of the 
           pre-$\beta''$ structure (circles) as a function of solute concentration. 
           Diamonds indicate the total energies obtained in direct DFT calculations for 
           the same structures.}
  \label{fig:toten2}
  \end{center}
\end{figure}

\figref{fig:toten1} shows that structures with an Mg:Si composition close 
to 1:1 are energetically favoured. 
This is in line with most experimental data on clusters and early precipitates 
in Al-Mg-Si alloys.\cite{Matsuda98} For this composition, the Matsuda phase is lowest in
energy, as found also in \refref{RaviWolv04}.
In order to picture how ``pre-$\beta''$-like'' structures 
may be build up, we also extracted the energetically most stable structures contained 
in the pre-$\beta''$ cell as a 
function of solute concentration $c_{\rm sol}\equiv c_{\rm Mg}+c_{\rm Si}=1/11,...,11/11$. Selecting 
the structures from the $2^{11}$ or the $3^{11}$ possible structures resulted 
in the same ground state structures. The energies are plotted in \figref{fig:toten2}   
together with directly calculated energies for the same structures. 
Thus, by comparing the two sets of energies, one gets an estimate of the predictive power 
of the CE method used. We note the the relaxation of the structure at $c_{\rm sol}=9/11$ did
not finish during the simulation. This could indicate that the structure is unstable.
Some representative lowest-energy structures in \figref{fig:toten2} are shown in Appendix B. 

%%%%%%%%%%%%%%%%%%%%%%%%%%%%%% FIGURE  4 %%%%%%%%%%%%%%%%%%%%%%%%%%%%%%%%%%%%%%%%%%%%%%%%%%%%%%%%%
%%%%%%%%%%%%%%%%%%%%%%%%%%%%%% end FIGURE 4 %%%%%%%%%%%%%%%%%%%%%%%%%%%%%%%%%%%%%%%%%%%%%%%%%%%%%%

The interface energies between secondary phase particles and a parent phase are 
important parameters in various theories of nucleation and phase stability 
that are usually hard to estimate. 
We calculated the pre-$\beta''$/Al and Matsuda/Al interface energies (at $T=0$) using 
Eq.~\eqref{eq:IsingSum}, see Table~\ref{tb:2}. 
The finding that the pre-$\beta''$/Al $({\bar 3}\; 1\; 0)$ interface is negative was
confirmed in a direct DFT calculation in a 44 atom super-cell. 
The result indicates that this phase may contain Al. 

The energy $E(\fsigma)$ of Eq.~\eqref{eq:CEenergy} is the static energy at constant volume. 
To this comes a lowering of elastic energy, $\Delta E^{\rm elast}$, due to volume and cell-shape relaxation 
of these phases. 
For the structures that were used in the fitting of the CE parameters, this term can be estimated
in the following way. Assuming that the phase $B$ relaxes centro-symmetrically within an $A$ matrix, which is 
assumed to be elastically isotropic, 
$\Delta E^{\rm elast}$ is given by\cite{Christian02}
\begin{equation}
  \label{eq:el-energy}
\Delta E^{\rm elast} = 18 \mu_A K_B/(3K_B+4\mu_A) \epsilon ^2 v_B
\end{equation}
Here, $\mu_A$ is the shear modulus of the $A$ phase and $K_B$ is the bulk modulus of the $B$ phase,
where in this estimate we used $K_{\rm Si,fcc}=K_{\rm Al}$. $v_B$ is 
the volume of $B$ phase and $\epsilon$ is the initial linear misfit of the $B$ phase relative to the 
$A$ phase. $\epsilon$ is in turn calculated from the internal pressure $p$ of the different structures
using $V_0 = V/(1-p/K)$. The results for some structures in Fig. \ref{fig:toten2} are given in 
Table~\ref{tb:3}.

%%%%%%%%%%%%%%%%%%%%%%%%%%%%%%%%%%%%%%%%%%%%%%%%%%%%%%%%%%%%%%%%%%%%%%%%%%%%%%%%%%%%%%%%
%%%%%%%%%%%%%%%%%%%%%%%%%%%%%%%%%%%%%%%%%%%%%%%%%%%%%%%%%%%%%%%%%%%%%%%%%%%%%%%%%%%%%%%% 
%%%%%%%%%                        DYNAMICA STABILITY                          %%%%%%%%%%%
%%%%%%%%%%%%%%%%%%%%%%%%%%%%%%%%%%%%%%%%%%%%%%%%%%%%%%%%%%%%%%%%%%%%%%%%%%%%%%%%%%%%%%%%
%%%%%%%%%%%%%%%%%%%%%%%%%%%%%%%%%%%%%%%%%%%%%%%%%%%%%%%%%%%%%%%%%%%%%%%%%%%%%%%%%%%%%%%%
\subsection{Dynamical stability of the Matsuda and pre-$\beta''$ phases}
The dynamic stability of the Matsuda phase and the pre-$\beta''$ phase was 
tested by displacing the atoms from their relaxed positions in the internally relaxed structures. 
For the Matsuda phase, starting with a 32 atom cell, we displaced the atoms 
in random directions $1/50$ of a nearest-neighbor distance. 
A subsequent relaxation showed that the forces were non-conservative 
(i.e., directed away from the initial positions) for the Si atoms within 
the Si plane.

A similar analysis of the pre-$\beta''$ structure shows that it is most likely 
metastable. In a separate calculation, we displaced the 
$(0,0,0)$ Mg atom in Fig.~\ref{fig:GP}a by a small distance in the z-direction. 
Also in this case, the force on the Mg atom was conservative.

In the case if the Matsuda phase, we note that, according to the findings in 
\refref{Matsuda98}, the structure exists in thin (one or a few atomic) $(0\;1\;0)$ 
slabs in the Al lattice (see Fig.~\ref{fig:GP}b). For such morphologies, the Matsuda phase will be stabilized 
by the surrounding Al lattice.

%%%%%%%%%%%%%%%%%%%%%%%%%%%%%%%%%%%%%%%%%%%%%%%%%%%%%%%%%%%%%%%%%%%%%%%%%%%%%%%%%%%%%%%%
%%%%%%%%%%%%%%%%%%%%%%%%%%%%%%%%%%%%%%%%%%%%%%%%%%%%%%%%%%%%%%%%%%%%%%%%%%%%%%%%%%%%%%%% 
%%%%%%%%%                        MONTE CARLO                          %%%%%%%%%%%
%%%%%%%%%%%%%%%%%%%%%%%%%%%%%%%%%%%%%%%%%%%%%%%%%%%%%%%%%%%%%%%%%%%%%%%%%%%%%%%%%%%%%%%%
%%%%%%%%%%%%%%%%%%%%%%%%%%%%%%%%%%%%%%%%%%%%%%%%%%%%%%%%%%%%%%%%%%%%%%%%%%%%%%%%%%%%%%%%
\subsection{Free energy of precipitation and kinetics of clustering}
In order to study precipitation and clustering it is necessary to know the solubility
limit of the alloying elements. In this sub-section we first describe the calculation of the solvus boundary 
in the Al rich end of the phase diagram, and then we present kMC simulations of clustering in the disordered
phase.

\begin{figure}
  \begin{center}
  \includegraphics*[width=8cm]{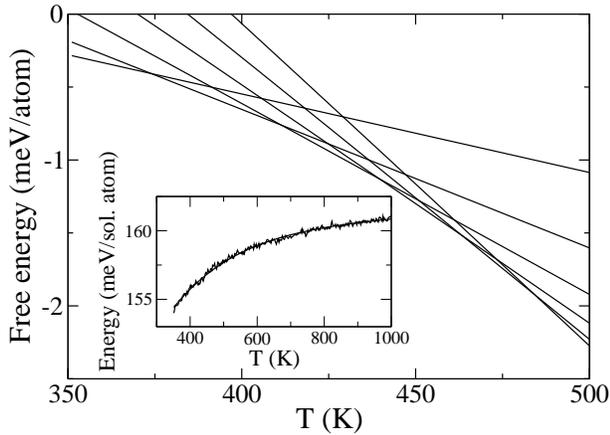}
  \caption{Free energies for the disordered phase, calculated by thermodynamic 
     integration from a reference state at 3000 K. The lines correspond to $c_{\rm sol}=1,2,...,6$ 
at.\% solutes,
     in order of increasing slope. The zero energy correspond to the reference state of completely
     separated L1$_0$ and Al phases, $E=c_{\rm sol}E_{\rm Al}+(1-c_{\rm sol})E_{\rm L1_0}$. The inset shows 
     one of the switching runs, at $c_{\rm sol} = 2\%$. The polynomial fit is included, and overlaps 
     with the energy points.}
  \label{fig:freeE}
  \end{center}
\end{figure}

\begin{figure}
  \begin{center}
  \includegraphics*[width=6cm]{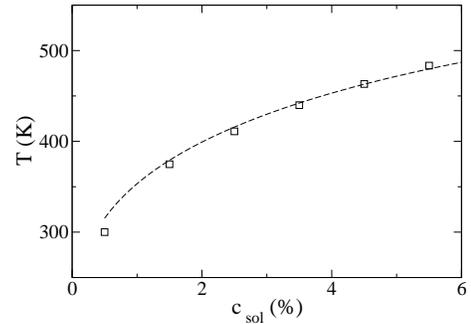}
  \caption{Calculated solvus line dividing the disordered and the two-phase regions of the phase 
    diagram for which $c_{\rm Mg}=c_{\rm Si}=c_{\rm sol}/2$ (squares). The dashed line corresponds
to an estimate based on the regular solution model.}
  \label{fig:Solvus}
  \end{center}
\end{figure}

We have focused on the subset of the phase diagram where 
the Mg:Si ratio is 1:1 and $c_{\rm sol} = c_{\rm Mg}+c_{\rm Si}$ varies between $0$ and $5\;\%$.
The energy per atom as a function of $T$ was calculated
for a range of 
concentrations, while staying in the disorderes phase.
In these simulations, an $80\times80\times80$ atom system was cooled at a rate of $0.02\;\K/$kMC step.
Based on a polynomial fit to the energy, we then used 
thermodynamic integration 
to calculate the entropy by use of Eqs.~\eqref{eq:ds_dh} and~\eqref{eq:S_HT}. The corresponding 
free energies are plotted in~\figref{fig:freeE}. The free energies are given with respect to the
normal state, which corresponds to separated Al and L$1_0$ phases at $0$~K. The solubility limit at each 
temperature corresponds to the lowest lying free-energy curve. We have implicitly assumed that the ordered phase,
the L$1_0$, is completely ordered, i.e., does not contain anti-defects or Al atoms. We checked the 
validity of this assumption, again using the regular solution model. It turns out that defects 
(typically 1-2\% below 500 K),
can be neglected in the calculation of the solubility limit at the Al rich end of the phase diagram. 
The resulting solvus-line is plotted in~\figref{fig:Solvus}. 

%%%%%%%%%%%%%%%%%%%%%%%%%%%%%% FIGURE 5 %%%%%%%%%%%%%%%%%%%%%%%%%%%%%%%%%%%%%%%%%%%%%%%%%%%%%%%%%
%%%%%%%%%%%%%%%%%%%%%%%%%%%%%% end FIGURE 5 %%%%%%%%%%%%%%%%%%%%%%%%%%%%%%%%%%%%%%%%%%%%%%%%%%%%%%%%%

%%%%%%%%%%%%%%%%%%%%%%%%%%%%%% FIGURE 6 %%%%%%%%%%%%%%%%%%%%%%%%%%%%%%%%%%%%%%%%%%%%%%%%%%%%%%%%%
%%%%%%%%%%%%%%%%%%%%%%%%%%%%%% end FIGURE 6 %%%%%%%%%%%%%%%%%%%%%%%%%%%%%%%%%%%%%%%%%%%%%%%%%%%%%%%%%

\begin{figure}
  \begin{center}
  \includegraphics*[width=7cm]{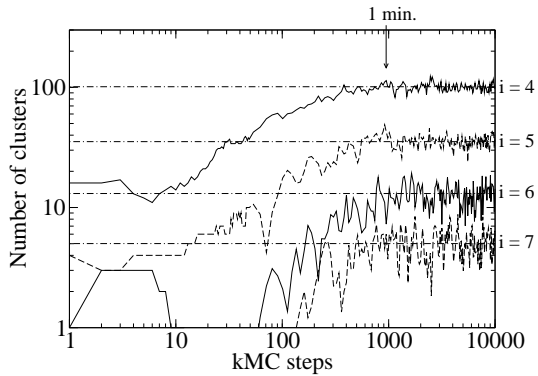}
  \caption{Log-log plot of the number of clusters of a given size $i$, calculated in a
kinetic Monte Carlo simulation of an $80\times80\times80$ atom system. The horizontal lines 
are the mean number of clusters over a total of $6\times10^6$ kMC steps.
kMC steps have been
related to physical time as described in the text.}
  \label{fig:Cluster}
  \end{center}
\end{figure}

Clustering in Al-Mg-Si was studied by kMC. 
The atoms were initially randomly distributed, corresponding to a previous homogenisation at a 
high temperature.
In \figref{fig:Cluster} we have displayed the cluster size distributions as a function of 
annealing time
for a 2\% solution at 400 K. The time-scale was calculated by assuming the Al self-diffusion rate
$D_{\rm Al} = 1.5\times10^{-5}\exp(-1.33/\kB/400)$ and a $10000$-fold increase in the number of vacancies,
corresponding to a homogenisation at 750 K. We believe that the latter factor is the dominant 
source of error in estimating the time-scale in this type of simulations. 
It is seen that the initial stage of clustering is fast ($\sim 1$ min.). In practice, since the 
kinetics is even faster at higher $T$, part of the initial clustering is expected to take place
already during the quench.

Since the system is not within the two-phase region, there are no nucleation and growth stages. 
After the initial clustering, the cluster sizes fluctuate in accordance with\cite{SoisMart00}
\begin{equation}
\label{eq:ClDist}
N^{\rm eq}(i)/N = \exp(-\Delta F(i)/\kB T)
\end{equation}
where $\Delta F(i)$ is the excess free energy required to form a cluster of size $i$, and
$N^{\rm eq}$ is the number of such clusters in a system containing $N$ atoms. $\Delta F(i)$
can be written as
\begin{equation}
\label{eq:ClEnergy}
\Delta F(i) = i\Delta f_{\rm vol}+Bi^{2/3}\sigma
\end{equation}
where $\Delta f_{\rm vol}$ is the volume free energy per atom of a cluster, $\sigma$ is the 
cluster/matrix interface free energy and $B$ is a geometrical factor. 

A longer kMC simulation ($6\times10^6$ kMC steps for $80\times80\times80$ atoms)  
showed that the cluster sizes were distributed in accordance with 
Eq.~\eqref{eq:ClDist} for clusters 
from $i=8$ up to $i = 16$, with an effective volume energy close to $0$ and an interface 
energy $\sigma=3\;{\rm meV/{\AA}}^2$, i.e., considerably lower than 
the mean interface energy of the Matsuda phase at $0\;\K$ (Table~\ref{tb:2}).
%%%%%%%%%%%%%%%%%%%%%%%%%%%%%% FIGURE 7 %%%%%%%%%%%%%%%%%%%%%%%%%%%%%%%%%%%%%%%%%%%%%%%
%%%%%%%%%%%%%%%%%%%%%%%%%%%%%% end FIGURE 7 %%%%%%%%%%%%%%%%%%%%%%%%%%%%%%%%%%%%%%%%%%%%

%%%%%%%%%%%%%%%%%%%%%%%%%%%%%%%%%%%%%%%%%%%%%%%%%%%%%%%%%%%%%%%%%%%%%%%%%%%%%%%%%%%%%%%%
%%%%%%%%%%%%%%%%%%%%%%%%%%%%%%%%%%%%%%%%%%%%%%%%%%%%%%%%%%%%%%%%%%%%%%%%%%%%%%%%%%%%%%%% 
%%%%%%%%%                         DISCUSSION	                             %%%%%%%%%%%
%%%%%%%%%%%%%%%%%%%%%%%%%%%%%%%%%%%%%%%%%%%%%%%%%%%%%%%%%%%%%%%%%%%%%%%%%%%%%%%%%%%%%%%%
%%%%%%%%%%%%%%%%%%%%%%%%%%%%%%%%%%%%%%%%%%%%%%%%%%%%%%%%%%%%%%%%%%%%%%%%%%%%%%%%%%%%%%%%

\section{Discussion}
The main purpose of this paper was to apply the CE-SIM method to model clusters 
and GP-zones in the Al-Mg-Si system. 
The interaction parameters $V_{\alpha,s}$ were fitted to reproduce the static energy 
$E$ of fcc based structures in the low solute-concentration regime. 
By use of cross-validation and by selecting clusters of increasing range, we 
ensure that an 'optimal' set of clusters is selected in the fitting procedure. 
Explicit calculations on 10 predicted ground-state structures derived from the 
pre-$\beta''$ structure reveal that the method holds its promise; the mean square 
deviation between predicted and directly calculated energies is $7.25$ meV/atom, 
to be compared with the cv-score $cv=7.1$ meV/atom in the fitting procedure. 

However, the energy of interest for a given $\fsigma$ is the Gibbs free energy 
$G=E-TS+pV$. 
The $pV$ term can be handled within the CE formalism,\cite{LaksFerr92} but it is much 
more difficult to treat the $TS$ term in a completely general manner. 
Our explicit calculations for the Matsuda phase show 
that it is dynamically unstable, i.e., the $TS$ term is undefined. 
In general, one expects that when the local configuration approaches an unstable 
structure, the vibrational frequencies are lowered (phonon softening), so that 
the structure is stabilized. 
Thus, this is an effect that we expect to be important in the early stages of 
precipitation in Al-Mg-Si. 
We note that in Al-Sc, the inclusion of the $TS$ term changes the phase diagram 
significantly.\cite{OzolAsta01} This is related to phonon softening but not to instabilities.
Instabilities in substitutional alloys have been studied previously, e.g. in 
W-Re.\cite{EkmaPers00_2} 

We now discuss the implications of our findings in relation to what is known 
from experiments on the nucleation of GP-zones and the 
$\beta''$-phase. 
In our simulations, we have only found one stable precipitate structure; the 
Mg$_1$Si$_1$ L$1_0$ structure. 
It forms in approximately spherical precipitates below $T\approx400\;\K$ for 
the composition $c_{\rm Mg}+c_{\rm Si}=2\;\%$. Matsuda et~al. report the 
formation of Mg$_1$Si$_1$ L$1_0$ precipitates at $T=323\;\K$ for an alloy containing $1.6$ mass~\%
Mg$_2$Si. These precipitates are in the form of $(0\;1\;0)$ planes. 
The morphology can be related to the internal stress of the Matsuda structure, which 
favours platelets in the $(0\;1\;0)$  plane. 
Because the platelets extend only one or a few planes in the normal direction, 
the Matsuda phase will be stabilized by the surrounding Al matrix. 

The pre-$\beta''$ and $\beta''$ phases are found experimentally to form at $T=423...523\;\K$ 
in alloys containing about 1.25 at.\% Al and Mg.\cite{MariAnde05}
\figref{fig:Solvus} shows the calculated phase boundary for forming Matsuda particles. 
One finds that 
neither the Matsuda phase nor the pre-$\beta''$ phase would form with the current 
atomic interaction parameters, at the temperatures and solute concentration 
where $\beta''$ forms.
Elastic relaxations would lower the energies of both the impurities and the ordered phases,
and are not expected to change this conclusion (see further \refref{RaviWolv04}). 
It is seen in Fig.~\ref{fig:Solvus} that the regular solution model describes the solvus phase 
boundary well up to $c_{\rm sol}\sim 4\%$. For a Mg:Si composition of 1:1, the expression for the 
solvus line according to this model is $c\approx 2\exp(-\Delta E/\kB T)$, in the low concentration limit. 
$\Delta E$ is the ordering 
energy which can be found in Fig.~\ref{fig:toten1} by comparing 
the energy of, e.g., the Matsuda phase with a line connecting the impurity energies of Mg and Si.
Thus, one finds that the ordering energy required to stabilize a structure at the temperature
and concentration where GP-zones form ($\approx 450\;\K$ and $1.5\%$ repectively), 
is at least $0.2\;{\rm eV/atom}$. This is larger than any of the ordering energies 
in the current study or in \refref{RaviWolv04}.

Therefore, in order to explain the nucleation mechanism of GP-zones and the $\beta''$ phase, 
we suggest the following possibilities.
Either, vibrational entropy effects or vacancies stabilize an ordered pre-$\beta''$ based structure, 
which then transforms into the $\beta''$ structure, or the structural transition takes 
place in the disordered phase, in a Mg-Si rich cluster with the right local 
configuration. Recent TEM observations clearly indicate that the GP-zones are 
ordered, fcc-based, structures.\cite{MariAnde05} Thus, the former explanation seems 
more likely.
In either case, there is a substantial gain in energy connected to the transition to
the $\beta''$ structure. 

Finally, we remark that it is likely that quenched-in vacancies play a direct role in the 
nucleation of the $\beta''$ phase. 
As pointed out in \refref{Derlet02}, vacancies may faciliate a structural transition 
from the pre-$\beta''$ to the $\beta''$ phase simply by providing the extra space 
needed for the (0,0,0) Mg atom to move $1/2$ lattice vector in the $z$-direction. 
Including vacancies in the CE description is formally straight-forward, however, in practice 
it would increase the complexity in the CE and lead to other technical complications.

%%%%%%%%%%%%%%%%%%%%%%%%%%%%%%%%%%%%%%%%%%%%%%%%%%%%%%%%%%%%%%%%%%%%%%%%%%%%%%%%%%%%%%%
%%%%%%%%%%%%%%%%%%%%%%%%%%%%%%%%%%%%%%%%%%%%%%%%%%%%%%%%%%%%%%%%%%%%%%%%%%%%%%%%%%%%%%% 
%%%%%%%%% CONCLUSIONS                                                     %%%%%%%%%%%
%%%%%%%%%%%%%%%%%%%%%%%%%%%%%%%%%%%%%%%%%%%%%%%%%%%%%%%%%%%%%%%%%%%%%%%%%%%%%%%%%%%%%%%
%%%%%%%%%%%%%%%%%%%%%%%%%%%%%%%%%%%%%%%%%%%%%%%%%%%%%%%%%%%%%%%%%%%%%%%%%%%%%%%%%%%%%%%
\section{Conclusions}
We have investigated the formation of GP-zones in the Al-Mg-Si system using a CE of the 
total energy. DFT pseudo-potential calculations of 154 fcc-based pure, binary and ternary 
structures were used to parameterize the energy by the SIM. Which cluster functions to include 
was chosen according to the cross-validation criterion, with the intension to avoid over-fitting.
The method successfully reproduces the total energy of 10 structures based on the suggested
pre-$\beta''$ structure. Regarding the formation of GP-zones, the CE calculations shows that the 
Mg$_1$Si$_1$ L$1_0$ structure suggested by Matsuda {\em et al.} is the most energetically stable phase.
This is also confirmed in direct DFT calculations. Our results are in fair agreement with the
DFT calculations presented by Ravi and Wolverton,\cite{RaviWolv04} where one should compare 
with their {\it constrained} GGA results for pre-$\beta''$ structures.
Regarding the formation of the pre-$\beta''$ and $\beta''$ structures,
we outline two possible nucleation mechanisms. Either, the vibrational entropy ($TS$) term,
or the presence of vacancies serve to stabilize the structures preceding the $\beta''$ phase, 
which then transforms to the meta-stable $\beta''$ phase. 

\section{Acknowledgments}
We thank M.~Ekman and C.~Wolverton for helpful discussions. 
M.~Ekman also kindly provided us with geometrical 
data for the CE. 
C.~D.~Marioara and S.~J.~Andersen contributed with their valuable knowledge about
precipitation in the Al-Mg-Si system. This work was funded through the Norwegian high performance 
computing programme and the Norwegian reseach Council through contract nr. 140553/I30.

%%%%%%%%%%%%%%%%%%%%%%%%%%%%%%%%%%%%%%%%%%%%%%%%%%%%%%%%%%%%%%%%%%%%%%%%%%%%%%%%%%%%%%%
%%%%%%%%%%%%%%%%%%%%%%%%%%%%%%%%%%%%%%%%%%%%%%%%%%%%%%%%%%%%%%%%%%%%%%%%%%%%%%%%%%%%%%% 
%%%%%%%%% BIBLIOGRAPHY                                                      %%%%%%%%%%%
%%%%%%%%%%%%%%%%%%%%%%%%%%%%%%%%%%%%%%%%%%%%%%%%%%%%%%%%%%%%%%%%%%%%%%%%%%%%%%%%%%%%%%%
%%%%%%%%%%%%%%%%%%%%%%%%%%%%%%%%%%%%%%%%%%%%%%%%%%%%%%%%%%%%%%%%%%%%%%%%%%%%%%%%%%%%%%%
%% \bibliography{refs}

%%%%%%%%%%%%%%%%%%%%%%%%%%%%%%%%%%%%%%%%%%%%%%%%%%%%%%%%%%%%%%%%%%%%%%%%%
% TABLE I                                                               %
%%%%%%%%%%%%%%%%%%%%%%%%%%%%%%%%%%%%%%%%%%%%%%%%%%%%%%%%%%%%%%%%%%%%%%%%%
% \twocolumn[\hsize\textwidth\columnwidth\hsize\csname
% @twocolumnfalse\endcsname
\begin{table}
% \widetext
\centering
\caption{Definition of cluster tested in the fitting procedure. In the final fit, we chose a small set
of cluster with a low $cv$-score. Each decoration listed, together with the corresponding cluster $\alpha$, makes up 
a distinct cluster function $\Phi_{\alpha,s}$.}
\label{tb:1}
\begin{tabular}{lll}
    & Cluster, $\alpha$       & Decorations, $s$                       \\
    & units of ($a_0/2$)                         &                     \\
\hline 
  Empty   &                   &                                        \\
  Point   & (000)             & 1; 2                                   \\
  P1      & (000) (110)       & 11; 12; 22                             \\
  P2      & (000) (200)       & 11; 12; 22                             \\
  ...      & ...              &                                        \\
  P7      & (000) (321)       & 11; 12; 22                             \\
  T1      & (000) (110) (101) & 111; 211; 222                          \\
  T2      & (000) (011) (0$\bar1$1) & 111; 211; 121; 221; 122; 222     \\
  T3      & (000) (110) ($\bar1$01) & 111; 211; 121; 221; 122; 222     \\
  T4      & (000) (121) (21$\bar1$) & 111; 211; 221; 222               \\
  T5      & (000) (211) (112)       & 111; 211; 121; 221; 122; 222     \\
  T6      & (000) (020) (112)       & 111; 211; 221; 112; 212; 222     \\
  T7      & (000) (110) (112)       & 111; 211; 121; 221;              \\
          &                         & 112; 212; 122; 222               \\
  Q1      & (000) (110) (101) (011) & 1111; 2111; 2211; 2221; 2222     \\
  Q2      & (000) (101) (011) (0$\bar1$1) & 1111; 2111; 2211; 1121; 2121;\\
          &                               & 2221; 1122; 2122; 2222     

\end{tabular}
\end{table}

\begin{table}
% \widetext
\centering
\caption{Interface energies between the Al phase and the Matsuda/pre-$\beta$ phase. Energies in 
meV/{\AA}$^2$. We checked the negative value for the Al/pre-$\beta''$ $(\bar{3}\;1\;0)$ interface in a direct
DFT calculation with result given in parenthesis.}
\label{tb:2}
\begin{tabular}{lll}
& Al/pre-$\beta''$ & \\
$(\bar{3}\;1\;0)$ & $(2\;3\;0)$ & $(0\;0\;1)$\\
\hline
$-3.57$ ($-4.31$) & $2.44$ & $6.44$\\
% \hline
% \hline
\\
& Al/Matsuda & \\
$(1\;0\;0)$ & $(0\;1\;0)$ & \\
\hline
$12.4$ & $5.9$ & 
\end{tabular}
\end{table}

\begin{table}
% \widetext
\centering
\caption{.}
\label{tb:3}
\begin{tabular}{l|ll}
Structure & pressure (meV/{\AA}$^3$) & Elastic energy \\
          &     (meV/{\AA}$^3$)      & (meV/B-phase atom)\\
\hline
4e (AlMgSi$_2$)     &  -2.5777 &    0.0488 \\
8c1 (AlMg$_3$Si$_4$)&   8.9378 &    0.5863\\
L1$_0$ (MgSi)         &   19.4610 &   2.7798\\
8c1 (AlMg$_4$Si$_3$)&   28.1928 &   5.8340\\
4a  (AlMg$_2$Si)    &   35.4267 &   9.2119\\
L1$_2$ (Al$_3$Mg)   &   32.1312 &   7.5778
\end{tabular}
\end{table}

%%%%%%%%%%%%%%%%%%%%%%%%%%%%%%%%%%%%%%%%%%%%%%%%%%%%%%%%%%%%%%%%%%%%%%%%%%%%%%%%%%%%%%%
%%%%%%%%%%%%%%%%%%%%%%%%%%%%%%%%%%%%%%%%%%%%%%%%%%%%%%%%%%%%%%%%%%%%%%%%%%%%%%%%%%%%%%% 
%%%%%%%%% APPENDIX                                                          %%%%%%%%%%%
%%%%%%%%%%%%%%%%%%%%%%%%%%%%%%%%%%%%%%%%%%%%%%%%%%%%%%%%%%%%%%%%%%%%%%%%%%%%%%%%%%%%%%%
%%%%%%%%%%%%%%%%%%%%%%%%%%%%%%%%%%%%%%%%%%%%%%%%%%%%%%%%%%%%%%%%%%%%%%%%%%%%%%%%%%%%%%%
\appendix
\section{Input structures}
%%%%%%%%%%%%%%%%%%%%%%%%%%%%%% FIGURE %%%%%%%%%%%%%%%%%%%%%%%%%%%%%%%%%%%%%%%%%%%%%%%%%%%%%%%%%
\begin{figure}
  \begin{center}
  %% {figs/Structures32.eps}
  \includegraphics*[width=6cm]{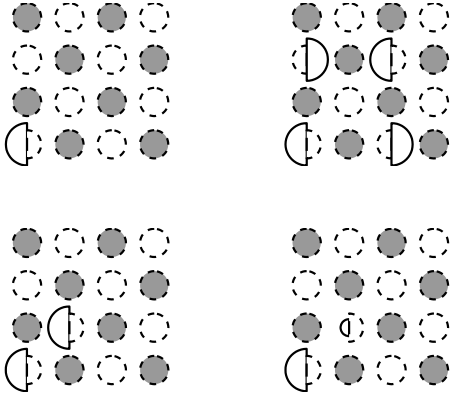}
  \caption{32 atom structures included in the fitting procedure. Dashed rings represent Al atoms, and 
  oversized/undersized rings represent Mg/Si atoms. New structures obtained by interchanging Mg
  and Si atoms were also included. Grey atoms are situated $a_0/2$ above the white atoms. Half-circles 
  indicate that two elements take up every second site in the vertical direction.}
  \label{fig:A32}
  \end{center}
\end{figure}

%%%%%%%%%%%%%%%%%%%%%%%%%%%%%% FIGURE %%%%%%%%%%%%%%%%%%%%%%%%%%%%%%%%%%%%%%%%%%%%%%%%%%%%%%%%%
\begin{figure}
  \begin{center}
  %% {figs/Structures16.eps}
  \includegraphics*[width=6cm]{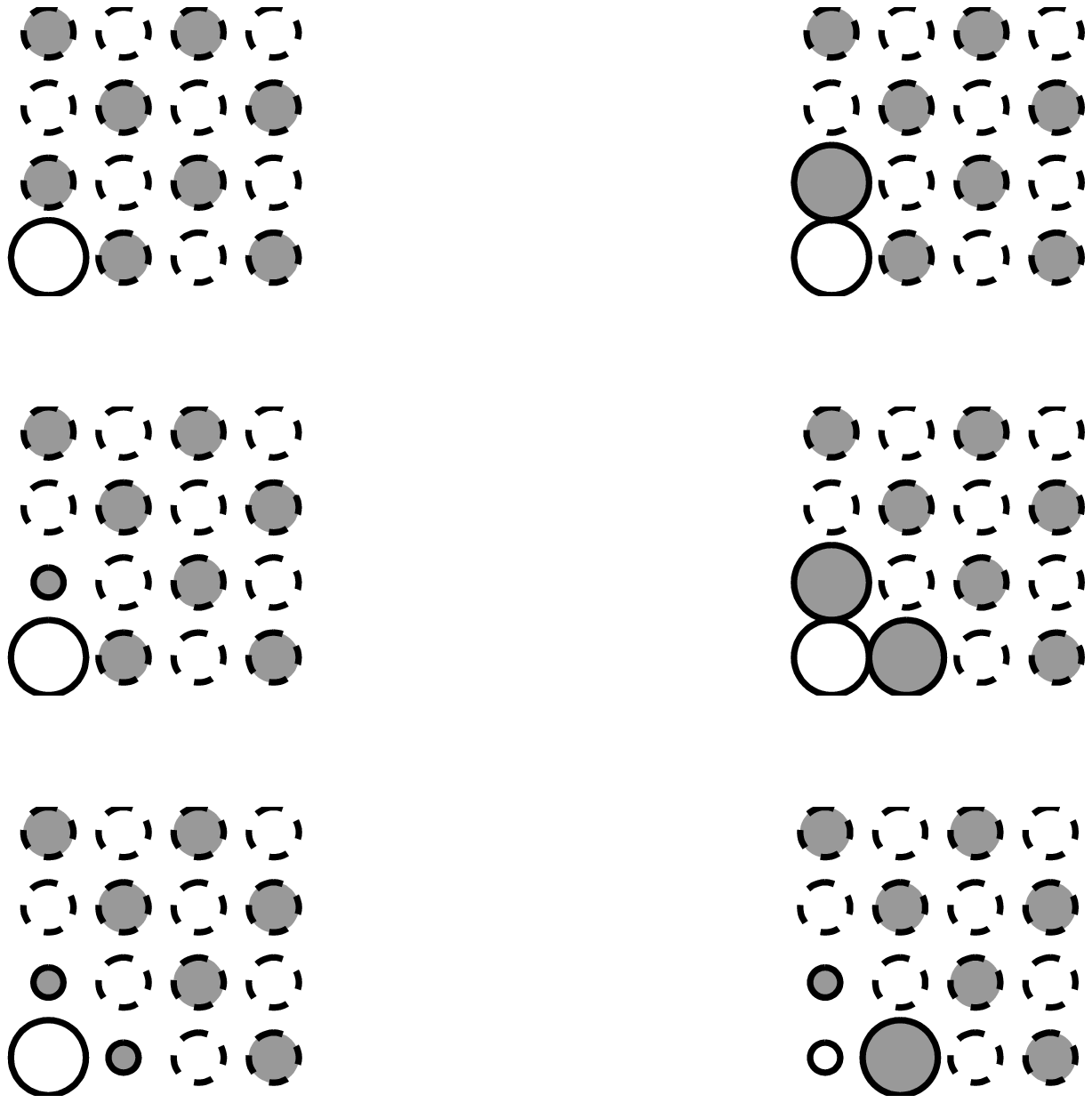}
  \caption{16 atom structures included in the fit. See further caption of Fig.~\ref{fig:A32}.}
  \label{fig:A16}
  \end{center}
\end{figure}

%%%%%%%%%%%%%%%%%%%%%%%%%%%%%% FIGURE %%%%%%%%%%%%%%%%%%%%%%%%%%%%%%%%%%%%%%%%%%%%%%%%%%%%%%%%%
\begin{figure}
  \begin{center}
  %% {figs/Structures12.eps}
  \includegraphics*[width=6cm]{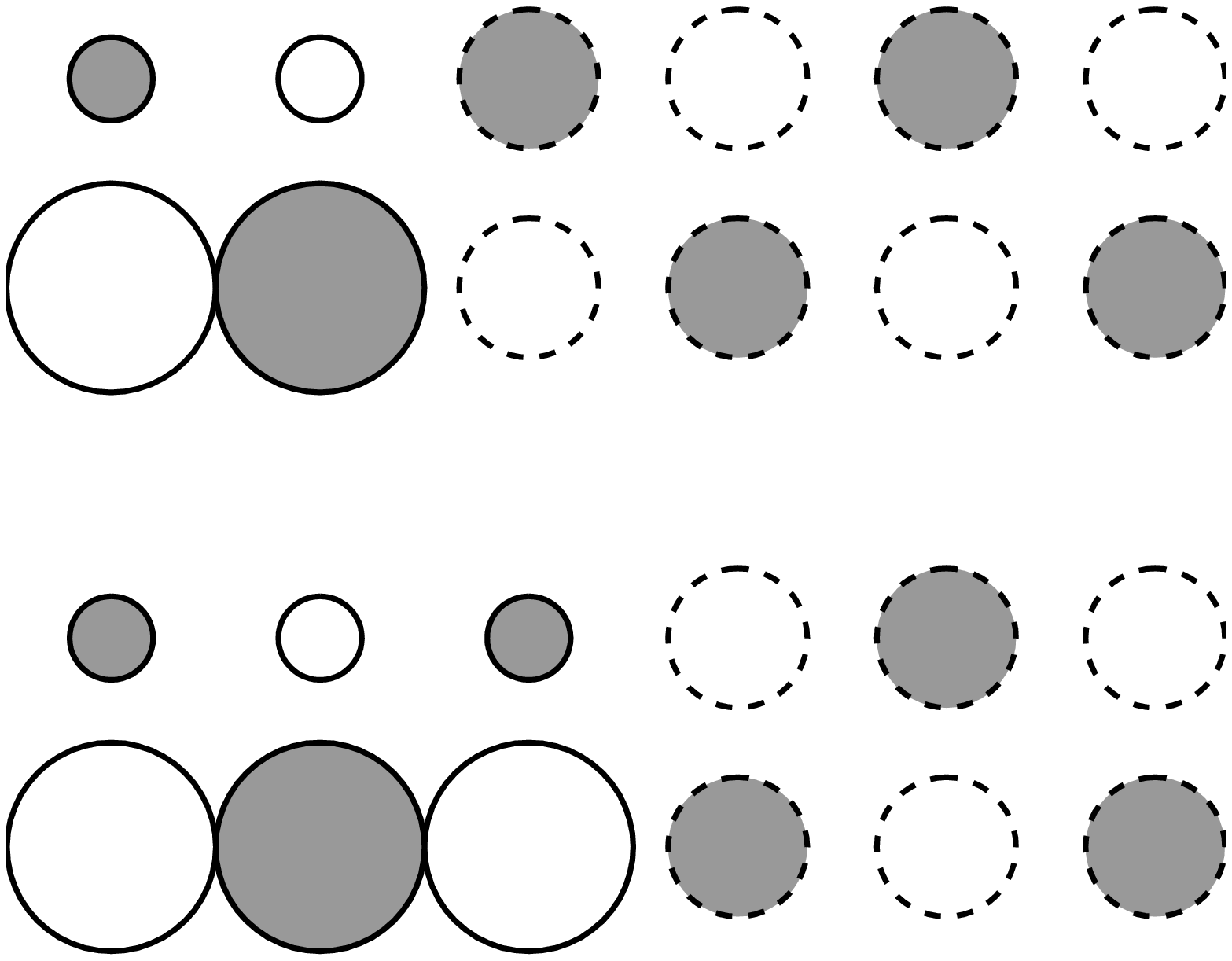}
  \caption{12 atom structures included in the fit. See further caption of Fig.~\ref{fig:A32}.}
  \label{fig:A12}
  \end{center}
\end{figure}

%%%%%%%%%%%%%%%%%%%%%%%%%%%%%% FIGURE %%%%%%%%%%%%%%%%%%%%%%%%%%%%%%%%%%%%%%%%%%%%%%%%%%%%%%%%%
\begin{figure}
  \begin{center}
  %% {figs/Structures8.eps}
  \includegraphics*[width=6cm]{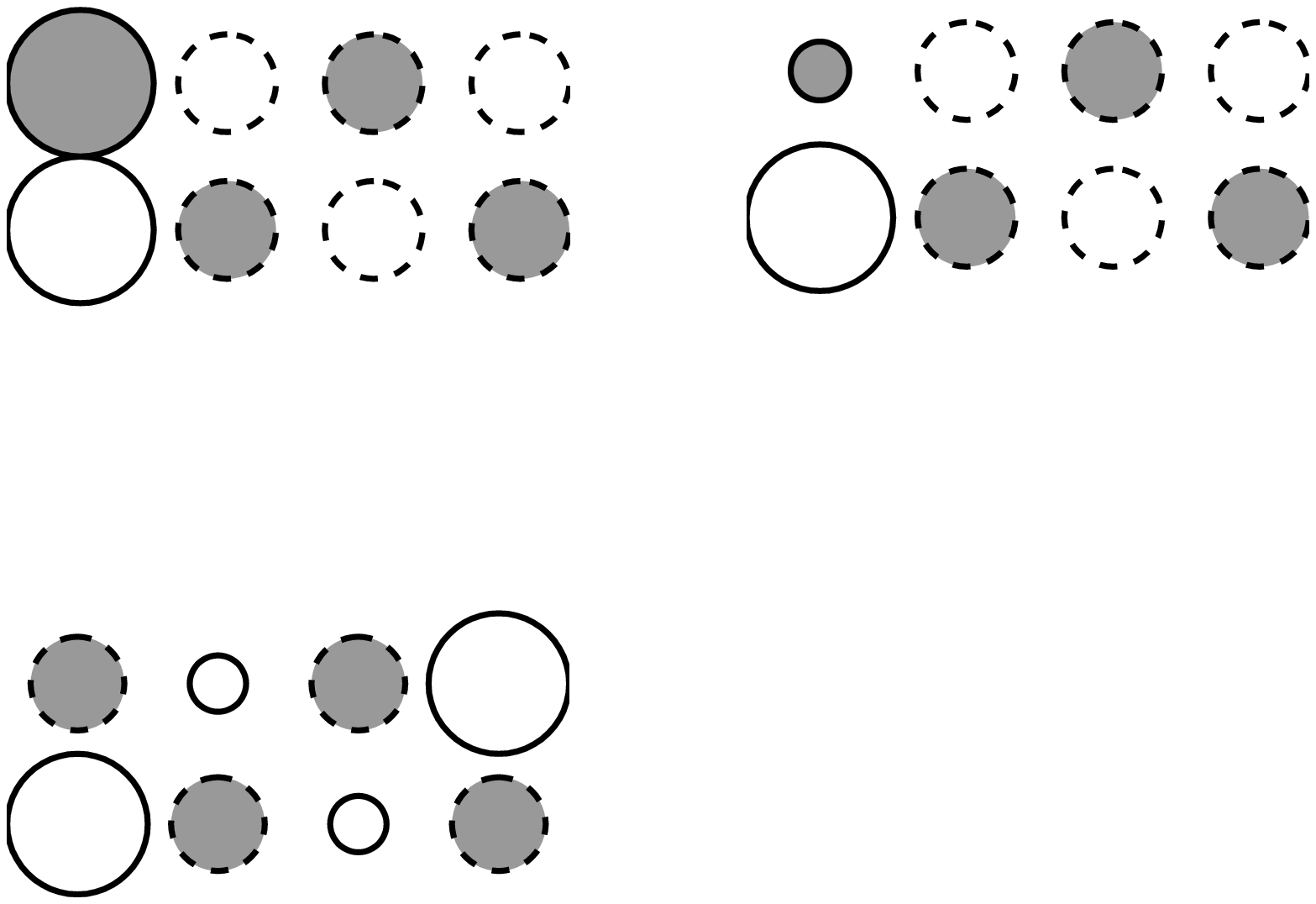}
  \caption{8 atom structures included in the fit. See further caption of Fig.~\ref{fig:A32}.}
  \label{fig:A8}
  \end{center}
\end{figure}

%%%%%%%%%%%%%%%%%%%%%%%%%%%%%%%%%%%%%%%%%%%%%%%%%%%%%%%%%%%%%%%%%%%%%%%%%%%%%%%%%%%%%%% 
%%%%%%%%% PRE-BETA LALALA                                                   %%%%%%%%%%%
%%%%%%%%%%%%%%%%%%%%%%%%%%%%%%%%%%%%%%%%%%%%%%%%%%%%%%%%%%%%%%%%%%%%%%%%%%%%%%%%%%%%%%%
\section{pre-$\beta''$-based structures}
% Some low-energy structures based on the pre-$\beta''$ structure. See further Fig.~\figref{fig:toten2}.
%%%%%%%%%%%%%%%%%%%%%%%%%%%%%% FIGURE %%%%%%%%%%%%%%%%%%%%%%%%%%%%%%%%%%%%%%%%%%%%%%%%%%%%%%%%%
% \begin{figure}
%   \begin{center}
%   \includegraphics*[width=6cm]{figs/fig1.eps}
%   \caption{Structures bla bla bla}
%   % \label{fig:ApreB}
%   \end{center}
% \end{figure}

% \begin{figure}
%   \begin{center}
%   \includegraphics*[width=6cm]{figs/fig2.eps}
%   \caption{2/11}
%   % \label{fig:ApreB}
%   \end{center}
% \end{figure}

\begin{figure}
  \begin{center}
  %% {figs/fig3.eps}
  \includegraphics*[width=12cm]{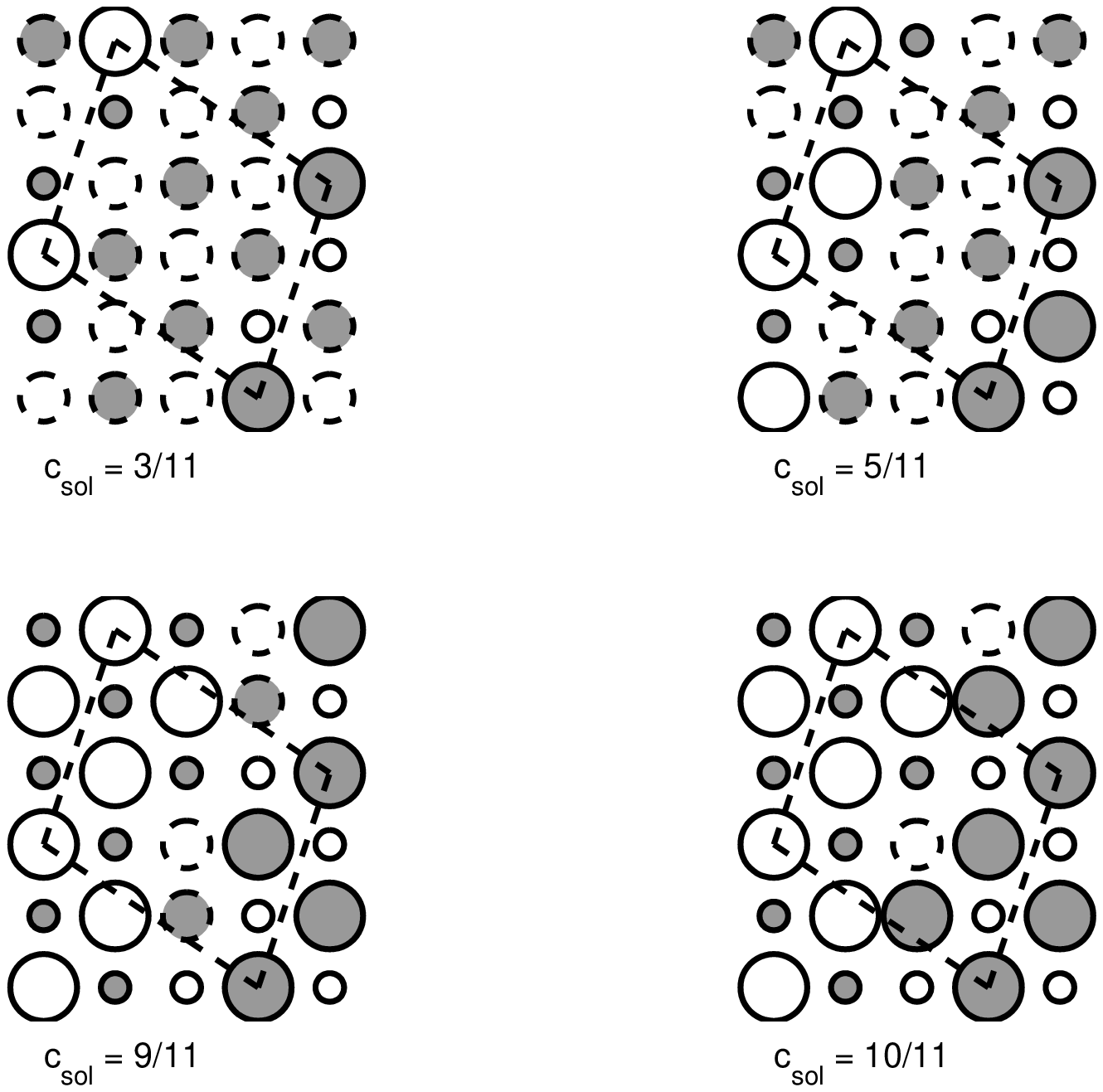}
  \caption{Some low-energy structures based on the pre-$\beta''$ structure. 
           See further \protect\figref{fig:toten2}.}
  % \label{fig:ApreB}
  \end{center}
\end{figure}

% \begin{figure}
%   \begin{center}
%   \includegraphics*[width=6cm]{figs/fig4.eps}
%   \caption{4/11}
%   % \label{fig:ApreB}
%   \end{center}
% \end{figure}

% \begin{figure}
%   \begin{center}
  %% \includegraphics*[width=6cm]{figs/fig5.eps}
%   \includegraphics*[width=4cm]{figure10.eps}
%   \caption{$c_{\rm sol}=5/11$}
%   \label{fig:5/11}
%   \end{center}
% \end{figure}

% \begin{figure}
%   \begin{center}
%   \includegraphics*[width=6cm]{figs/fig6.eps}
%   \caption{6/11}
%   % \label{fig:ApreB}
%   \end{center}
% \end{figure}

% \begin{figure}
%   \begin{center}
%   \includegraphics*[width=6cm]{figs/fig7.eps}
%   \caption{7/11}
%   % \label{fig:ApreB}
%   \end{center}
% \end{figure}

% \begin{figure}
%   \begin{center}
%   \includegraphics*[width=6cm]{figs/fig8.eps}
%   \caption{8/11}
%   % \label{ig:ApreB}
%   \end{center}
% \end{figure}

% \begin{figure}
%   \begin{center}
  %% \includegraphics*[width=6cm]{figs/fig9.eps}
%   \includegraphics*[width=4cm]{figure11.eps}
%   \caption{$c_{\rm sol}=9/11$}
%   \label{fig:9/11}
%   \end{center}
% \end{figure}

% \begin{figure}
%   \begin{center}
%   %% \includegraphics*[width=6cm]{figs/fig10.eps}
%   \includegraphics*[width=4cm]{figure12.eps}
%   \caption{$c_{\rm sol}=10/11$}
%   \label{fig:10/11}
%   \end{center}
% \end{figure}

\end{document}